\documentclass[aps,prd,reprint,groupedaddress, showpacs, floatfix]{revtex4-1}

\usepackage{times}
\usepackage{euscript}
\usepackage{amsthm}
\usepackage{graphicx}
\usepackage{amsmath}
\usepackage{amssymb}
\usepackage{amsfonts}
\usepackage[english]{babel}
\usepackage{epsfig}
\usepackage{multirow,makecell,array}
\usepackage{mathtools}
\usepackage{color}

\begin{document}


\title{Momentum distribution of particles created\\in space-time-dependent colliding laser pulses}


\author{I.~A.~Aleksandrov$^{1, 2}$}\author{G.~Plunien$^{3}$} \author{V.~M.~Shabaev$^{1}$}
\affiliation{$^1$~Department of Physics, St. Petersburg State University, 7/9 Universitetskaya Naberezhnaya, Saint Petersburg 199034, Russia\\$^2$~ITMO University, Kronverkskii Avenue 49, Saint Petersburg 197101, Russia\\$^3$~Institut f\"ur Theoretische Physik, Technische Universit\"at Dresden, Mommsenstrasse 13, Dresden D-01062, Germany
}


\begin{abstract}
We study the pair-production process in the presence of two counterpropagating linearly polarized short laser pulses. By means of a nonperturbative technique, we take into account the full coordinate dependence of the external field going beyond the dipole and standing-wave approximations. In particular, we analyze the momentum distribution of particles created. It is demonstrated that the spatial variations of the laser pulses may play a crucial role. The more accurate treatment reveals a number of prominent features: the pair-production probabilities become considerably smaller, the quantitative behavior of the momentum spectra changes dramatically, and the pulse shape effects become much less pronounced. The results of our study are expected to be very important for future theoretical and experimental investigations.
\end{abstract}

\maketitle
\section{Introduction}\label{sec:intro}
Electron-positron pair production (PP) in strong external fields is a fundamental nonlinear phenomenon predicted by quantum electrodynamics~\cite{sauter_1931, euler_heisenberg, schwinger_1951}. In the presence of a quasistatic electric background, the vacuum decay rate becomes non-negligible if the corresponding field strength approaches the Schwinger critical value $E_\text{c} = m^2c^3/(|e|\hbar) = 1.3 \times 10^{16}$~V/cm ($m$ and $e$ are the electron mass and charge, respectively), which is not accessible to modern experimental facilities. However, because of the recent progress in laser technology, it is expected that, in the near future, one will be able to achieve the peak electric field strength $E_0$ which is sufficiently close to $E_\text{c}$~\cite{dipiazza_rmp_2012} (e.g., ELI-Ultra High Field Facility~\cite{eli_web} aims to attain $E_0/E_\text{c} \sim 10^{-3}$). Moreover, the Schwinger limit effectively lowers in the presence of rapid oscillations of the external field. An oscillating background can be characterized by the dimensionless adiabaticity parameter $\xi = |eE_0|/(mc\omega)$, where $\omega$ is the corresponding frequency ($\xi$ is the inverse of the Keldysh parameter $\gamma$~\cite{keldysh}). If $\xi \ll 1$, the PP process can be accurately described by means of perturbation theory. Within this multiphoton regime, the phenomenon of pair creation has already been observed experimentally~\cite{burke_prl_1997}. In the opposite limit $\xi \gg 1$, i.e., the Schwinger (tunneling) regime, the process can only be considered with the aid of nonperturbative calculations. The corresponding studies are supposed not only to illuminate various theoretical aspects of the PP phenomenon but also to advance the search for most favorable experimental scenarios and thus to make the observation of the Schwinger effect feasible.

One of these scenarios could be a collision of two high-intensity laser pulses~\cite{brezin_itzykson, popov, nar-nik_1974, mostepanenko_1974} (see also~\cite{popov_2001, avetissian_pre_2002, dipiazza_prd_2004, bulanov_2006, hebenstreit_prl_2009, ruf_prl_2009, mocken_pra_2010} and references therein) which is the focus of our investigation. In this article we assume that the pulses have the same linear polarization along the $x$ axis: $\vec{E} \parallel \vec{e}_x$, $\vec{B} \parallel \vec{e}_y$, and $z$ is the propagation direction. Accordingly, an individual pulse can be described by the following vector potential (hereafter we employ the relativistic units $\hbar=c=1$): 
\begin{eqnarray}
&&A_y = A_z = 0, \notag \\
&&A^{(\pm)}_x (t, z) = -\frac{E_0}{\omega}F(\omega t \mp k_z z) \sin (\omega t \mp k_z z + \varphi),\label{eq:pulse_potential}
\end{eqnarray}
where $E_0$ is the peak electric field strength, $\omega$ is the carrier frequency ($k_z = \omega$), $\varphi$ describes the carrier-envelope phase (CEP), and $F(\eta)$ is a smooth envelope function. If the pulse contains several cycles, the spatial dependence of the carrier is stronger than that of the envelope. This means that one can neglect the latter: $F(\omega t \mp k_z z) \to F(\omega t)$. In this case, the resulting field $A_x (t, z) = A^{(+)}_x (t, z) + A^{(-)}_x (t, z)$ becomes a standing wave oscillating with time, and the vector potential reads
\begin{equation}
A^{(\text{SWA})}_x (t, z) = -\frac{2E_0}{\omega}F(\omega t) \sin (\omega t + \varphi) \cos (k_z z).\label{eq:potential_swa}
\end{equation}
In what follows, this approximation will be referred to as the standing-wave approximation (SWA). If one assumes that the $e^+ e^-$ pairs basically form in the vicinity of the points $z_n = \pi n/k_z$ ($n \in \mathbb{Z}$) where the electric field is maximal and the magnetic component vanishes, one can further simplify the expression~(\ref{eq:potential_swa}) by neglecting the coordinate dependence of the carrier:
\begin{equation}
A^{(\text{DA)}}_x (t) = -\frac{2E_0}{\omega}F(\omega t) \sin (\omega t + \varphi).\label{eq:potential_da}
\end{equation}
The external field is now purely electric and spatially homogeneous. We will call this the dipole approximation (DA). It is worth noting that DA is expected to provide adequate predictions only when the laser wavelength is much larger than the typical length scale of the pair-production process $2mc^2/|eE_0|$. This requirement is equivalent to the condition $\xi \gg 1/\pi$.

Most theoretical studies of the problem have been conducted within DA~\cite{brezin_itzykson, popov, nar-nik_1974, mostepanenko_1974, popov_2001, avetissian_pre_2002, dipiazza_prd_2004, bulanov_2006, hebenstreit_prl_2009, mocken_pra_2010, abdukerim_plb_2013, aleksandrov_prd_2017, fillion_pra_2012, dumlu_prd_2010, li_prd_2015}. These investigations identified the main general patterns of the PP process and revealed a number of distinctive features regarding the pulse shape, governed by the envelope function $F(\eta)$ and the CEP parameter $\varphi$~\cite{dipiazza_prd_2004, hebenstreit_prl_2009, abdukerim_plb_2013, aleksandrov_prd_2017}. Taking into account the coordinate dependence of the external field is a very challenging task. The effects of the spatial inhomogeneities of the laser pulses were partially included in Refs.~\cite{woellert_prd_2015, woellert_plb_2016} where the role of the laser pulse polarization was examined within SWA. To our knowledge, the only study beyond SWA was reported in Ref.~\cite{ruf_prl_2009} where the resonant Rabi oscillations and the resonant behavior of the PP probabilities were discussed. It was shown that the influence of the spatial variations of the external field and its magnetic component may play a very important role in the PP process. However, the momentum distribution of particles created was not investigated.

Within the present study we go beyond DA and SWA taking into account the spatial dependence of both the carrier and the envelope. We demonstrate that these approximations fail to correctly describe the momentum spectrum of particles. As will be shown below, a number of characteristic features revealed within DA and SWA arise merely due to the inaccuracy of the corresponding approximations. Moreover, both DA and SWA substantially overestimate the PP probabilities.

In order to analyze the pulse shape effects, we vary the parameter $\varphi$ and employ two different envelope functions. First, we use a ``flat'' profile which has an extended plateau region:
\begin{equation}
F_1 (\eta)=
\begin{cases}
\sin^2 \big [ \frac{1}{2} (\pi N - |\eta|) \big ] &\text{if}~~\pi (N-1) \leq |\eta| < \pi N,\\
1 &\text{if}~~|\eta| < \pi (N-1),\\
0 &\text{otherwise},
\end{cases} \label{eq:F_flat}
\end{equation}
where $N$ is the number of cycles in the pulse, so the pulse duration is $\tau = 2 \pi N/\omega$. Second, we employ a slowly varying $\cos^2$ envelope:
\begin{equation}
F_2 (\eta)= \cos^2 \bigg ( \frac{\eta}{2N} \bigg ) \theta (\pi N - |\eta|).
\label{eq:F_cos2}
\end{equation}
In this study we choose $E_0 = 0.1 E_\text{c}$ and $\tau = 2\times 10^{-19}~\text{s}$ and use $N = 1-5$. This corresponds to $\omega/m \approx 0.04-0.20$ and $2 \xi \approx 1.0-4.9$. The adiabaticity parameter $2 \xi$ regards the resulting field of the two laser pulses. This choice of the laser background parameters ensures that the classical-external-field approximation is valid and the PP process reflects nonperturbative nature. We examine ultrashort laser pulses since it allows us to clearly illustrate the main findings of our investigation and helps us to save the computational time, which becomes extremely important beyond SWA.

In order to evaluate the necessary PP probabilities, we utilize the nonperturbative technique described in Ref.~\cite{aleksandrov_prd_2016}. The method is based on the general formalism~\cite{fradkin_gitman_shvartsman} of quantization in the Furry picture and is briefly described in Appendix~\ref{sec:appendix_1}. To attest our numerical procedures, we first reproduced the main results of Refs.~\cite{hebenstreit_prl_2009, ruf_prl_2009, mocken_pra_2010, woellert_prd_2015}. Note that the potential~(\ref{eq:pulse_potential}) does not vanish at $|t| \to +\infty$. To make sure that our computations provide the probabilities of the production of real particles (electrons and positrons), one has to introduce a temporal window function $R(t)$, so the resulting vector potential reads
\begin{equation}
A_x (t, z) = \big [ A^{(+)}_x (t, z) + A^{(-)}_x (t, z) \big ] R(t).\label{eq:potential_total}
\end{equation}
The function $R(t)$ is a smooth function which slowly vanishes when the pulses do not overlap. If this profile is sufficiently wide, our results converge and do not depend on the shape of this function (for more details, see Appendix~\ref{sec:appendix_2}). In Fig.~\ref{fig:scheme} the process is illustrated in the $t-z$ plane.
\begin{figure}[t]
\center{\includegraphics[width=1.0\linewidth]{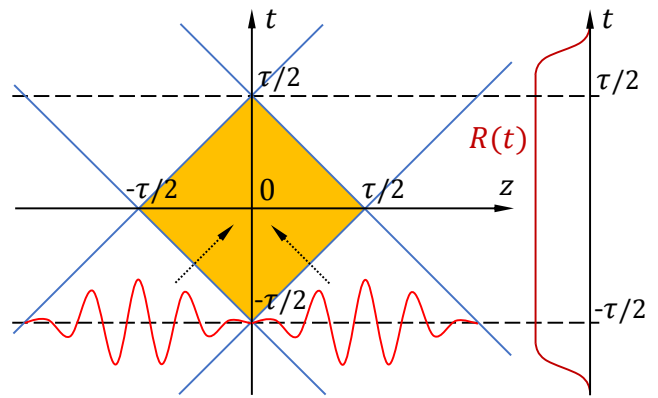}}
\caption{The dynamics of two counterpropagating laser pulses in the $t-z$ plane. The pulses overlap within the square region (yellow). The switching function $R(t)$ has a wide plateau region and slowly vanishes outside this interval.}
\label{fig:scheme}
\end{figure}
%
\section{Dipole and standing-wave approximations}\label{sec:da_swa}
\begin{figure}[t]
\center{\includegraphics[width=1.0\linewidth]{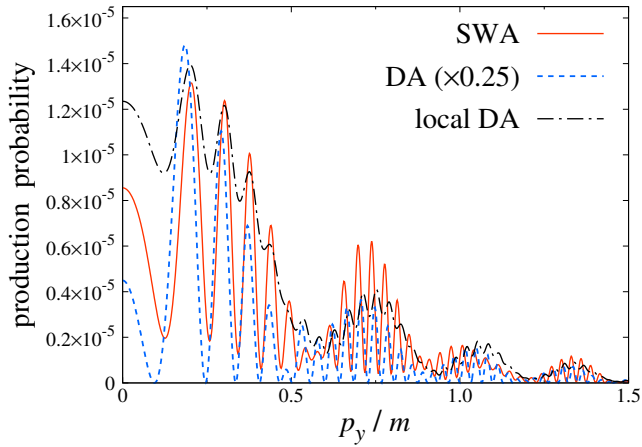}}
\caption{The momentum spectrum of particles created with $p_x=p_z=0$ in the case of the flat envelope calculated within DA, SWA, and local DA ($N=5$, $\varphi = \pi/2$). The results obtained in DA are multiplied by a factor of $0.25$.}
\label{fig:swa_vs_da_flat}
\end{figure}

Let us first examine the momentum spectra obtained within DA and SWA. In Fig.~\ref{fig:swa_vs_da_flat} they are depicted for the case of the flat envelope (\ref{eq:F_flat}) as a function of the momentum component along the magnetic field direction $y$. The values
represent the mean number of electrons (positrons) produced per unit volume (if the spin state is not taken into account, the results should be multiplied by a factor of $2$). Due to the symmetry of the vector potential, the spectrum is invariant with respect to the reflection $\boldsymbol{p} \to - \boldsymbol{p}$. A very important difference between DA and SWA arises due to the fact that, within the former, the external field is purely electric and thus the $y$ and $z$ axes are equivalent in DA. In contrast, the spectrum found within SWA behaves differently along these directions. It turns out that the magnetic component of the resulting field drastically alters the momentum distribution in the $p_x-p_z$ plane, so we compare the $p_y$ dependences fixing $p_x = p_z = 0$. In Fig.~\ref{fig:swa_vs_da_flat} we observe a quite nontrivial behavior which is reproduced within both DA and SWA, provided the results obtained in DA are multiplied by a factor of about $0.25$. This quantitative discrepancy can be accounted for if one
performs the calculations within DA for various $E_0 (z) = E_0 \cos(k_z z)$ and then averages the results over one cycle of the standing wave (we refer to this procedure as local DA). This leads to the spectrum which does not require any additional normalization (see Fig.~\ref{fig:swa_vs_da_flat}).

The oscillatory behavior in Fig.~\ref{fig:swa_vs_da_flat} contains two different scales. We observe slow oscillations with period $\Delta p \approx \omega$ and faster oscillations with period $\delta p \approx \omega/10$. The former were found within DA in Ref.~\cite{hebenstreit_prl_2009} (see also Refs.~\cite{aleksandrov_prd_2017, abdukerim_plb_2013, dumlu_prl_2010, fillion_pra_2012, akkermans_prl_2012}) for the case of a Gaussian envelope profile and interpreted as resonances in a one-dimensional quantum-mechanical scattering problem. We suppose that the faster oscillations appear for the similar reasons due to the nonmonochromaticity of the laser pulses (i.e., their Fourier transform contains modes with various frequencies depending on $N$). Since each pulse contains $N$ cycles of the carrier, the envelope function is essentially a half-cycle oscillation with frequency $\omega/N$. The analysis of pulses with other values of $N$ showed that indeed $\delta p \approx \omega/(2N)$. Note that within local DA the faster oscillations are suppressed. Although the averaging allows one to partially include the spatial inhomogeneities of the external field, this approach does not properly capture the effects of the interference among modes with various frequencies.

\begin{figure}[t]
\center{\includegraphics[width=1.0\linewidth]{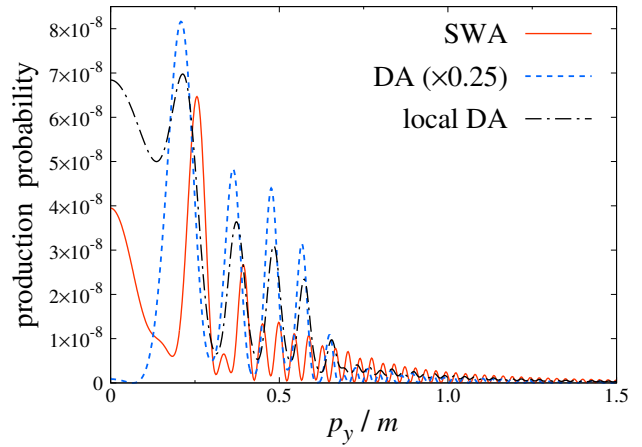}}
\caption{The momentum spectrum of particles created in the case of the $\cos^2$ envelope calculated within DA, SWA, and local DA ($N=5$, $\varphi = \pi/2$). The results obtained in DA are multiplied by a factor of $0.25$.}
\label{fig:swa_vs_da_cos2}
\end{figure}

The pulse shape effects become evident when we employ the $\cos^2$ envelope function (see Fig.~\ref{fig:swa_vs_da_cos2}). In this case, the slow oscillations disappear, which indicates that the interference  strongly depends on the pulse shape. Moreover, one observes that the production probability is now about $2$ orders of magnitude smaller than that presented in Fig.~\ref{fig:swa_vs_da_flat}. An obvious reason for this could be that the envelope functions (\ref{eq:F_flat}) and (\ref{eq:F_cos2}) satisfy $F_1(\eta) \geq F_2(\eta)$.

Finally, we study the effects of CEP. It turns out that this parameter plays a very important role. For instance, within SWA for $\varphi = 0$, the PP probabilities are much smaller than those found for $\varphi = \pi/2$ (the number of particles integrated over $p_y$ is $4.9$ times smaller). Despite this great quantitative difference, the two spectra have nearly the same qualitative behavior discussed above.

In the following, we will study how our findings should change if one goes beyond SWA and takes into account the spatial variations of both the carrier and the envelope.

\section{Beyond SWA}\label{sec:beyond_swa}
First, one has to note that taking into consideration the coordinate dependence of the envelope function makes the system finite in the $z$ direction. We normalize our results, so one obtains again the mean number of particles per unit volume (see Appendix~\ref{sec:appendix_1}). This allows one to
directly compare the spectra found beyond SWA to those presented in Figs.~\ref{fig:swa_vs_da_flat} and \ref{fig:swa_vs_da_cos2}.

\begin{figure}[t]
\center{\includegraphics[width=1.0\linewidth]{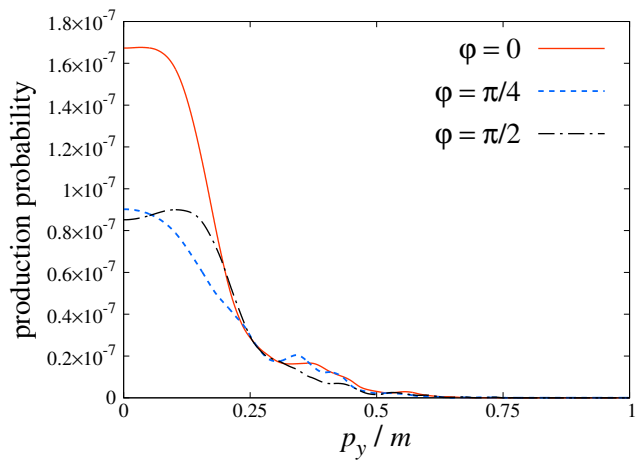}}
\caption{The momentum spectrum evaluated beyond SWA in the case of the flat envelope for various values of the CEP parameter $\varphi$ ($N=5$).}
\label{fig:clp_1}
\end{figure}
\begin{figure*}[t]
\center{\includegraphics[width=0.48\linewidth]{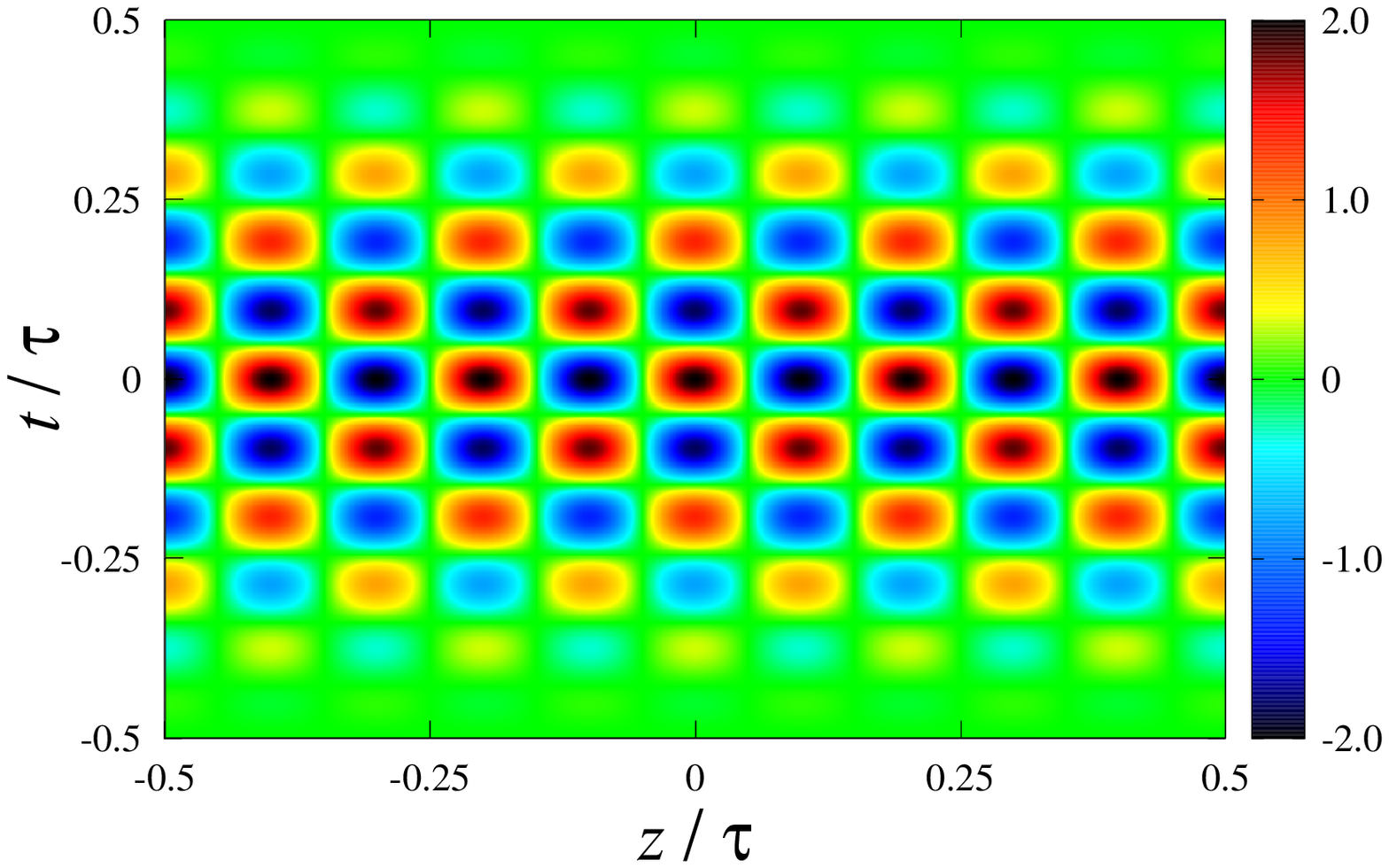}~~~
\includegraphics[width=0.48\linewidth]{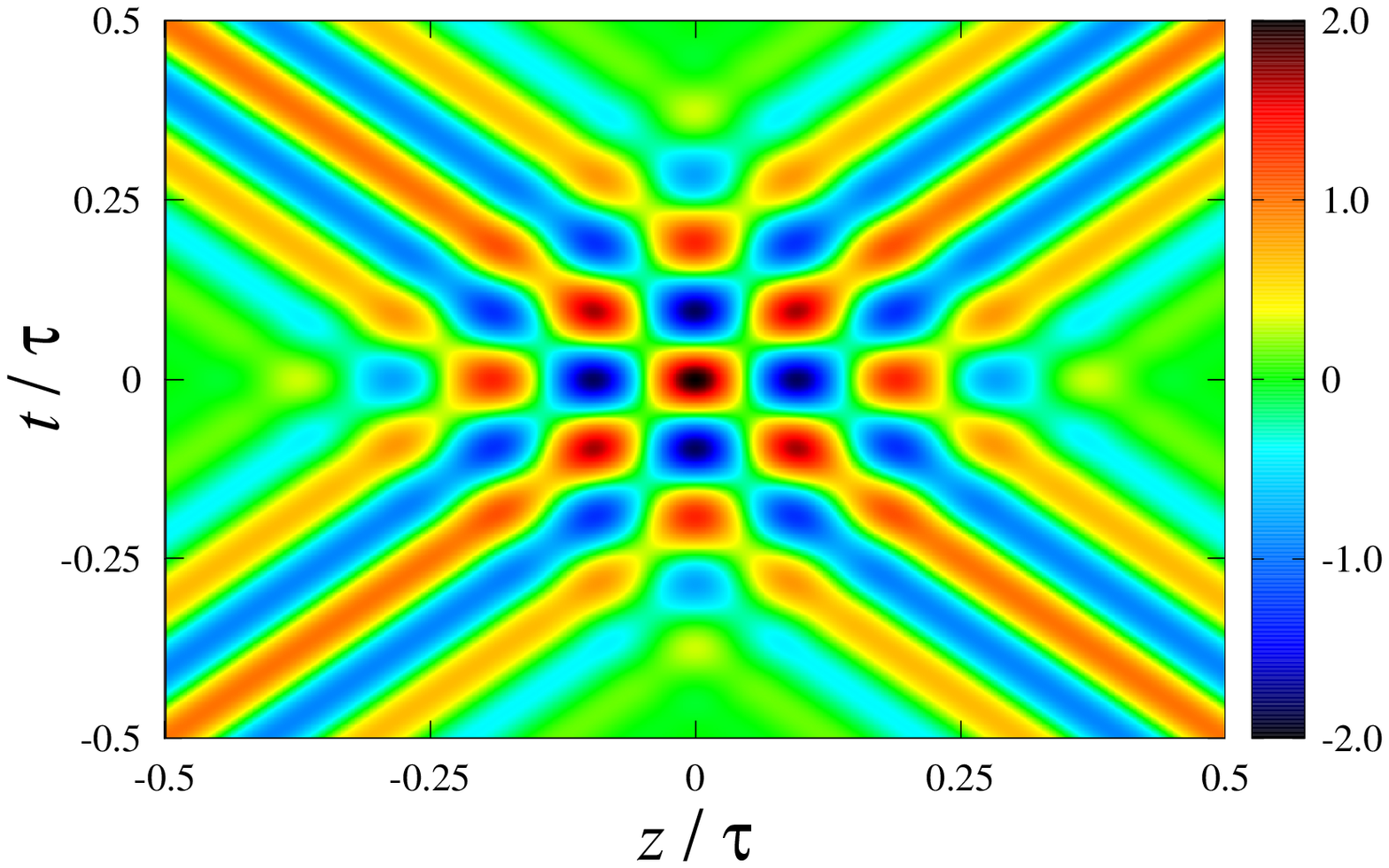}}
\caption{The electric field strength $E_x (t, z)$ (in units of $E_0$) calculated within SWA (left) and beyond SWA (right).}
\label{fig:field}
\end{figure*}
In Fig.~\ref{fig:clp_1} we depict the spectrum of particles produced for the case of the flat profile~(\ref{eq:F_flat}) and three different values of the CEP parameter $\varphi$ ($N=5$). First, one observes that the oscillatory structure now disappears. The reason for this is that the temporal dependence of the external field is not separated from the spatial one as in the previously used approximations. Whereas within DA and SWA, the time-dependent factor $F(\omega t)$ induces the oscillatory patterns discussed above, beyond SWA the temporal and spatial variations are related, and therefore the overall dynamics in the collision process does not exhibit any resonant behavior. Besides, the spectra are now much less sensitive to the CEP parameter. The quantitative discrepancy between the red curve ($\varphi = 0$) and the others in Fig.~\ref{fig:clp_1} is the largest difference that we observed within our study beyond SWA. Finally, we note that the PP probabilities evaluated beyond SWA are about $2$ orders of magnitude smaller. For instance, for $\varphi = 0$, the number density integrated over $p_y$ is now $3.66\times 10^{-8}$, while in SWA it amounts to $8.38\times 10^{-6}$. This can be understood if one notes that beyond SWA the overlap region occupies much less area in the $t-z$ plane. According to the normalization used (see Appendix~\ref{sec:appendix_1}), one should compare the field configurations within the ``unit'' interval $z \in [-\tau/2,~\tau/2]$.  In Fig.~\ref{fig:field} we display the electric field strength $E_x (t, z)$ within SWA and beyond this approximation. We observe that in SWA the field occupies the whole interval $z \in [-\tau/2,~\tau/2]$ and has the same amplitude for each value of $z$ depending only on time via the factor $F(\omega t)$. Beyond SWA, the overlap region is considerably smaller. Thus, the neglect of the spatial finiteness of the external pulses leads to a great overestimation of the particle yield.
\begin{figure}[t]
\center{\includegraphics[width=1.0\linewidth]{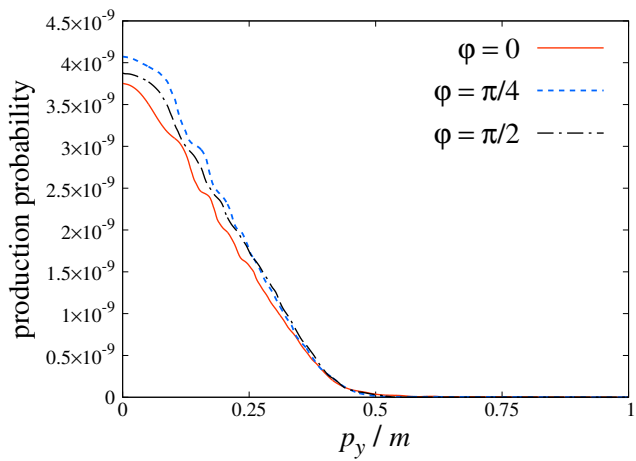}}
\caption{The momentum spectrum evaluated beyond SWA in the case of the $\cos^2$ envelope for various values of the CEP parameter $\varphi$ ($N=5$).}
\label{fig:clp_2}
\end{figure}

In Fig.~\ref{fig:clp_2} we present the momentum spectra of particles for the case of the $\cos^2$ envelope. We observe that the momentum distribution is almost independent of the CEP parameter in contrast to our findings within DA and SWA. As was expected, in the case of the flat profile, the PP probabilities are greater than those found for the $\cos^2$ envelope function. Nevertheless, the difference is now much smaller in comparison to SWA, which indicates again that the previously used approximations fail to provide adequate quantitative predictions. The qualitative behavior of the spectra in Fig.~\ref{fig:clp_2} does not strongly differ from that presented in Fig.~\ref{fig:clp_1}, and as in the case of the flat envelope, the spectra are now much narrower than they were within DA and SWA.
A crucial difference between the field configurations in DA and SWA and that arising beyond SWA relates to the way the external field is being switched on and off. Beyond SWA the envelope $F(\eta)$ governs the spatial profiles of the two colliding pulses while the function $R(t)$ is now responsible for switching on and off the external background. Once the results are stable with respect to the changes of the function $R(t)$, the physical quantities are not so sensitive to the parameters of the spatial envelope $F(\eta)$. On the other hand, the profile $F(\omega t)$ plays a different role within DA and SWA. Apart from the appearance of the oscillatory structure discussed above, this fact also leads to a number of striking effects regarding the pulse shape and carrier-envelope phase. Beyond DA and SWA these effects are suppressed. The analysis of pulses with other values of $N$ confirms the aforementioned findings.

\section{Conclusion}\label{sec:conclusion}
In the present study it was demonstrated that the dipole and the standing-wave approximations predict a number of well-pronounced effects that do not appear once more precise calculations extended to the (1+1)-dimensional laser field configuration are performed. In particular, it was found that the oscillatory patterns in the momentum spectrum of particles created vanish beyond SWA, the PP probabilities become much smaller, and the spectra are much less sensitive to the pulse shape parameters (in the recent studies~\cite{gies_prl_2016, gies_prd_2017}, it was also found that the imaginary part of the effective action in a spatially inhomogeneous background obeys universal scaling laws). This strongly suggests that DA and SWA do not properly describe the quantitative as well as qualitative characteristics of the momentum distribution of particles produced. The results obtained within these approximations should be treated and interpreted very carefully, for they may not reflect the real patterns. This point is extremely important for experimental studies since they could involve a great number of parameters which cannot be varied without any guidance. More accurate theoretical predictions are needed if one attempts to identify most promising experimental setups for the practical observation of the Schwinger effect. We note that several techniques for pulse shape optimization were developed for the case of spatially homogeneous backgrounds in Refs.~\cite{hebenstreit_plb_2014, fillion_arxiv_2017, kohlfuerst_prd_2013, hebenstreit_plb_2016}. Our findings indicate a great importance of the spatial variations of the external fields in the particular scenario of two counterpropagating short pulses. However, we expect that going beyond the dipole approximation is strongly required within a much broader class of problems concerning quantum dynamics in external fields (see also, e.g., Refs.~\cite{hebenstreit_prl_2011, kohlfuerst_plb_2016}). For instance, calculations similar to those performed in our study may play a significant role in the context of the dynamically assisted Schwinger effect~\cite{schutzhold_prl_2008} (see also Refs.~\cite{akal_prd_2014, otto_plb_2015, schneider_jhep_2016} and references therein). It is already known that the subcycle structure of the external pulses and the shape of their temporal profiles have a notable impact on the pair-creation probabilities~\cite{nuriman_plb_2012, kohlfuerst_prd_2013, hebenstreit_plb_2014, linder_prd_2015, torgrimsson_2017, hebenstreit_plb_2016}. Nevertheless, of particular interest is the role of the spatial inhomogeneities of the electromagnetic pulses. This is an important issue for future research.

Finally, we point out that the analysis of more realistic field configurations should also be important within the studies of the Schwinger mechanism in other fields of physics. Similar processes can be identified in various nanostructures (e.g., in graphene~\cite{allor_prd_2008, gavrilov_prd_2012, fillion_prb_2015}) and with ultracold atoms in optical lattices~\cite{szpak_njp_2012, kasper_plb_2016, kasper_njp_2017}. In quantum chromodynamics, quark-antiquark pairs can be produced in relativistic heavy ion collisions (see, e.g., Refs.~\cite{gelis_prl_2006, tanji_aop_2010, ruggieri_npa_2015}).
\section*{Acknowledgments}
This investigation was supported by RFBR (Grant No.~17-52-12049), by Saint Petersburg State University (Grants No.~11.42.660.2017, No.~11.42.666.2017, No.~11.65.41.2017, and No.~11.38.237.2015), and by DFG (Grants No.~PL~254/10-1 and No.~STO~346/5-1). I. A. A. acknowledges the support from the German-Russian Interdisciplinary Science Center (G-RISC) funded by the German Federal Foreign Office via the German Academic Exchange Service (DAAD), from TU Dresden (DAAD-Programm Ostpartnerschaften), and from the Basis foundation.
\appendix
\section{Nonperturbative technique}\label{sec:appendix_1}
According to the general formalism described in detail in Ref.~\cite{fradkin_gitman_shvartsman}, one can extract all the information about the PP probabilities from two special sets of time-dependent solutions of the Dirac equation. Assuming that the external field vanishes for $t \leq t_\text{in}$ and for $t \geq t_\text{out}$, we define these solutions by the following conditions:
\begin{equation}
{}_\zeta \Psi_n (t_\text{in}, \boldsymbol{x}) = {}_\zeta \Psi^{(0)}_n (\boldsymbol{x}),\quad {}^\zeta \Psi_n (t_\text{out}, \boldsymbol{x}) = {}^\zeta \Psi^{(0)}_n (\boldsymbol{x}),\label{eq:psi_in_out}
\end{equation}
where ${}_\zeta \Psi^{(0)}_n (\boldsymbol{x})$ and ${}^\zeta \Psi^{(0)}_n (\boldsymbol{x})$ are the eigenfunctions of the Dirac Hamiltonian considered at $t=t_\text{in}$ and $t=t_\text{out}$, respectively, and $\zeta$ is the sign of the corresponding energy eigenvalues. The sets $\{{}_\zeta \Psi_n \}$ and $\{{}^\zeta \Psi_n \}$ are orthonormal and complete. One can rigorously demonstrate~\cite{fradkin_gitman_shvartsman} that the mean number of electrons (positrons) produced with given quantum numbers $m$ can be evaluated as
\begin{eqnarray}
n^-_m &=& \sum_n G({}^+|{}_-)_{mn} G({}_-|{}^+)_{nm}, \label{eq:num_el}\\
n^+_m &=& \sum_n G({}^-|{}_+)_{mn} G({}_+|{}^-)_{nm}, \label{eq:num_pos}
\end{eqnarray}
where the $G$ matrices can be defined as the inner products
\begin{eqnarray}
G({}^\zeta|{}_\kappa)_{nm} &=& ({}^\zeta \Psi_n,~{}_\kappa \Psi_m), \label{eq:G_inner_product_1}\\
G({}_\zeta|{}^\kappa)_{nm} &=& ({}_\zeta \Psi_n,~{}^\kappa \Psi_m). \label{eq:G_inner_product_2}
\end{eqnarray}
We first consider the problem beyond SWA, i.e., employ Eqs.~(\ref{eq:pulse_potential}) and (\ref{eq:potential_total}). Since the asymptotic solutions in~(\ref{eq:psi_in_out}) are essentially plane waves (in what follows, $m = \{ \boldsymbol{p}, r \}$ where $r$ determines a spin state), it is convenient to work in the momentum representation. Besides, the transverse momentum $\boldsymbol{p}_\perp = (p_x,~p_y)$ is conserved. Then, for instance, the out solution with $\zeta = +$ reads
\begin{equation}
{}^+ \Psi_{\boldsymbol{p}, r} (t, \boldsymbol{x}) = \frac{\mathrm{e}^{i\boldsymbol{p}_\perp \boldsymbol{x}_\perp}}{(2\pi)^{3/2}} \int \limits_{-\infty}^{+\infty} dk \, \mathrm{e}^{i (p_z - k)z}\,{}^+g_{\boldsymbol{p}, r} (t, k). \label{eq:fourier_def}
\end{equation}
The function ${}^+g_{\boldsymbol{p}, r}$ obeys
\begin{equation}
{}^+g_{\boldsymbol{p}, r} (t_\text{out}, k) = u_{\boldsymbol{p}, r} \delta (k), \label{eq:g_initial}
\end{equation}
where $p_0 = \sqrt{\boldsymbol{p}^2 + m^2}$ and $u_{\boldsymbol{p}, r}$ is a constant bispinor corresponding to the positive-energy states ($u^\dagger_{\boldsymbol{p}, r} u_{\boldsymbol{p}, r'} = \delta_{rr'}$). The Dirac equation in terms of the function ${}^+g_{\boldsymbol{p}, r}$ reads
\begin{widetext}
\begin{equation}
i \partial_t \,{}^+g_{\boldsymbol{p}, r} (t, k) = \boldsymbol{\alpha}_\perp \, \boldsymbol{p}_\perp \,{}^+g_{\boldsymbol{p}, r} (t, k) + \beta m \,{}^+g_{\boldsymbol{p}, r} (t, k) + (p_z - k) \alpha_z {}^+g_{\boldsymbol{p}, r} (t, k) - e \alpha_x \int \limits_{-\infty}^{+\infty} dq \, a_x (t, k - q) \,{}^+g_{\boldsymbol{p}, r} (t, q), \label{eq:g_dirac2}
\end{equation}
\end{widetext}
where $a_x (t, k)$ is the Fourier transform of the vector potential. Using the ``initial'' condition~(\ref{eq:g_initial}), we propagate the function ${}^+g_{\boldsymbol{p}, r}$ backwards in time according to Eq.~(\ref{eq:g_dirac2}). The step of the $k$ grid should be sufficiently small, so the delta function in Eq.~(\ref{eq:g_initial}) is represented properly. The specific forms~(\ref{eq:F_flat}) and~(\ref{eq:F_cos2}) of the envelope function are particularly convenient for our computations since the Fourier transform of the vector potential can be found analytically.

The matrix $G({}_-|{}^+)_{nm}$ can be evaluated as follows:
\begin{equation}
G({}_-|{}^+)_{\boldsymbol{p},r;\, \boldsymbol{p}',r'}=\delta(\boldsymbol{p}_\perp+\boldsymbol{p}'_\perp)v^\dagger_{\boldsymbol{p},r}\, {}^+ g_{\boldsymbol{p}', r'} (t, p_z + p'_z),
\label{eq:Gg}
\end{equation}
where the bispinor $v_{\boldsymbol{p},r}$ relates to the negative-energy states. By means of Eqs.~(\ref{eq:num_el}) and~(\ref{eq:Gg}), we find the mean number of particles created per unit cross-section area:
\begin{equation}
\frac{(2\pi)^2}{V_{xy}} \frac{dN_{\boldsymbol{p},r}}{d^3 \boldsymbol{p}} = \sum_{r'=\pm 1} \int \limits_{-\infty}^{+\infty} dk \, |v^\dagger_{-\boldsymbol{p}_\perp, k,r'} {}^+g_{\boldsymbol{p}, r} (t, p_z+k)|^2.
\label{eq:number}
\end{equation}
Here we employ a conventional regularization $(2\pi)^2 \delta(\boldsymbol{p}_\perp = 0) = V_{xy}$. This approach allows one to directly evaluate the PP probabilities by propagating the PP amplitudes themselves rather than the individual one-particle solutions in the coordinate space. We solve Eq.~(\ref{eq:g_dirac2}) for each momentum of the particle produced in parallel.

Within SWA the calculations become much less complicated since the spatial dependence of the vector potential contains only $\cos (k_z z)$ which is a sum of two plane waves. Accordingly, the problem naturally becomes discrete and the integral in Eq.~(\ref{eq:g_dirac2}) is reduced to the sum of two terms corresponding to $k+k_z$ and $k-k_z$ (see also Refs.~\cite{woellert_prd_2015, aleksandrov_prd_2016}). The numerical procedure now yields a number of particles per unit volume as $2\pi \delta(p_z = 0) = L_z$. To compare the spectra evaluated beyond SWA with those found in SWA, we multiply the former by $2\pi/\tau = \omega/N$ since $\tau$ represents a characteristic size of the system in the $z$ direction. Within DA the component $p_z$ is also conserved, so the problem becomes one dimensional, and it is necessary to solve an ordinary differential equation, which can be done even more efficiently.
\section{Temporal window function $R(t)$}\label{sec:appendix_2}
\begin{figure}[ht]
\center{\includegraphics[width=1.0\linewidth]{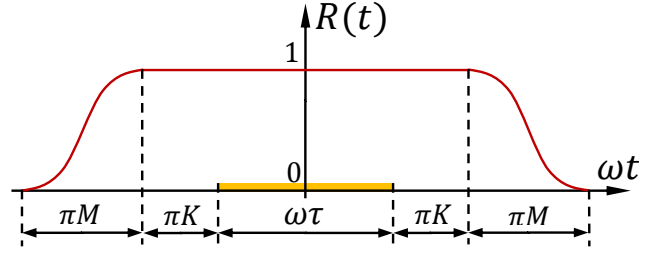}}
\caption{The temporal window function $R(t)$. The yellow bar denotes the overlap interval.}
\label{fig:R_scheme}
\end{figure}

The function $R(t)$ should be introduced in order to properly define the in and out solutions and employ the theoretical formalism described above. It is clear that the main contribution to the pair-production probabilities arises due to the overlap of the two laser pulses since an individual plane wave (i.e., single laser pulse) does not produce $e^+e^-$ pairs. The function $R(t)$ has a sufficiently wide plateau which is $K$ cycles broader than the overlap interval (see Fig.~\ref{fig:R_scheme}). The switching parts of this function contain $M$ half-cycles each.
It turns out that once the function $R(t)$ is introduced, our numerical procedure yields nonzero results for the case of an individual pulse. This nonphysical contribution should be subtracted when one evaluates the number density $n (\boldsymbol{p})$ of particles created by the superposition of two colliding pulses:
\begin{equation}
n (\boldsymbol{p}) = n^{(\text{II})} (\boldsymbol{p}) - n^{(+)} (\boldsymbol{p}) - n^{(-)} (\boldsymbol{p}),
\label{eq:subtraction}
\end{equation}
where $n^{(\text{II})} (\boldsymbol{p})$ is the number density obtained numerically for the case of two pulses \big[i.e., employing Eq.~(\ref{eq:potential_total})\big] and $n^{(\pm)} (\boldsymbol{p})$ are the results calculated for the right- and left-propagating pulses, respectively. This subtraction leads to reliable results if the following conditions are satisfied:
\begin{enumerate}
\item The values of the function $n (\boldsymbol{p})$ converge as one increases $K$ and $M$. As $K,~M \to \infty$, the pulses are being switched on and off when they are far from each other and therefore generate pairs independently.
\item The results do not depend on the shape of the switching parts of the function $R(t)$.
\end{enumerate}

\begin{figure}[ht]
\center{\includegraphics[width=1.0\linewidth]{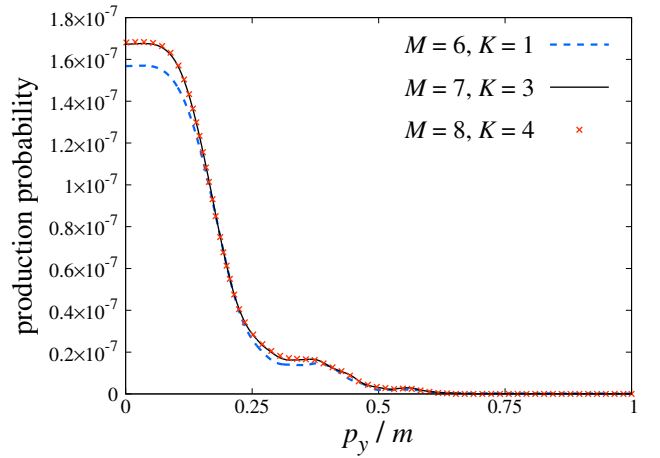}}
\caption{The momentum spectrum $n (\boldsymbol{p})$ evaluated beyond SWA for $p_x=p_z=0$ and various values of the parameters $K$ and $M$ of the temporal window function $R(t)$ ($N=5$, $\varphi=0$).}
\label{fig:conv}
\end{figure}

In our calculations we confirmed both of these conditions. For instance, in Fig.~\ref{fig:conv} the momentum distribution of particles for the case of the flat envelope is presented for various values of the pair $K$, $M$ ($N=5$, $\varphi=0$). As these parameters increase, the results converge. We performed our calculations for various shapes of the switching parts and obtained identical results, although the individual terms in Eq.~(\ref{eq:subtraction}) were rather sensitive to these changes. This sensitivity resembles that regarding the pulse shape effects within DA and SWA.

\begin{figure}[ht]
\center{\includegraphics[width=1.0\linewidth]{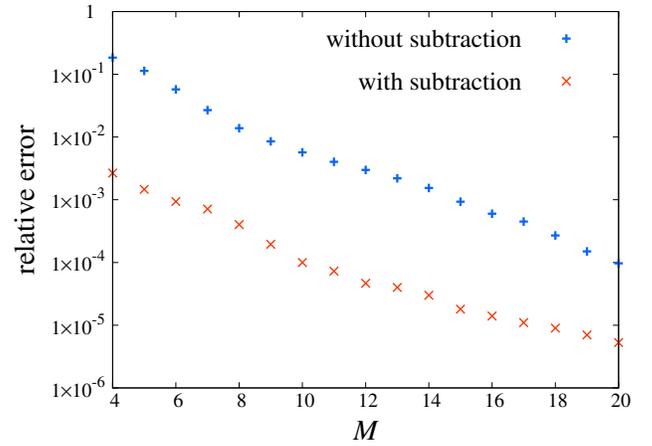}}
\caption{The relative error as a function of $M$ for the number density of particles produced with $\boldsymbol{p}=0$ ($K=4$).}
\label{fig:conv_M}
\end{figure}

It turns out that as one increases the duration of the switching parts of the profile $R(t)$ ($M \to \infty$), the contributions $n^{(\pm)} (\boldsymbol{p})$ continually decrease, which means that, in principle, one can perform the computations for very large $M$ without any subtraction procedure. However, this drastically increases the computational time, while the subtraction~(\ref{eq:subtraction}) makes the convergence much faster. In Fig.~\ref{fig:conv_M} the corresponding relative errors are depicted as a function of $M$. Although the results converge to the same limit, the accuracy is about 1-2 orders of magnitude higher if one employs the subtraction procedure. Besides, it is reasonable to choose the shape of the function $R(t)$ so that the nonphysical contributions are as small as possible. This allows one to perform more accurate and efficient calculations. We found that smooth and slowly varying profiles are more advantageous in this respect.
%



\begin{thebibliography}{99}
%
\bibitem{sauter_1931} F.~Sauter, Z. Phys. {\bf 69}, 742 (1931).
%
\bibitem{euler_heisenberg} W.~Heisenberg and H.~Euler, Z. Phys. {\bf 98}, 714 (1936).
%
\bibitem{schwinger_1951} J.~Schwinger, Phys. Rev. {\bf 82}, 664 (1951).
%
\bibitem{dipiazza_rmp_2012} A.~Di Piazza, C.~M\"uller, K.Z. Hatsagortsyan, and C.H. Keitel, Rev. Mod. Phys. {\bf 84}, 1177 (2012).
%
\bibitem{eli_web} https://eli-laser.eu/.
%
\bibitem{keldysh} L.~V.~Keldysh, Zh. Eksp. Teor. Fiz. {\bf 47}, 1945 (1964) [Sov. Phys. JETP {\bf 20}, 1307 (1965)].
%
\bibitem{burke_prl_1997} D.~L.~Burke, R.~C.~Field, G.~Horton-Smith, J.~E.~Spencer, D.~Walz, S.~C.~Berridge, W.~M.~Bugg, K.~Shmakov, A.~W.~Weidemann, C.~Bula, K.~T.~McDonald, E.~J.~Prebys, C.~Bamber, S.~J.~Boege, T.~Koffas, T.~Kotseroglou, A.~C.~Melissinos, D.~D.~Meyerhofer, D.~A.~Reis, and W.~Ragg, Phys. Rev. Lett. {\bf 79}, 1626 (1997).
%
\bibitem{brezin_itzykson} E.~Brezin and C.~Itzykson, Phys. Rev.~D {\bf 2}, 1191 (1970).
%
\bibitem{popov} V.~S.~Popov, Pis'ma Zh. Eksp. Teor. Fiz. {\bf 13}, 261 (1971) [JETP Lett. {\bf 13}, 185 (1971)]; Zh. Eksp. Teor. Fiz. {\bf 61}, 1334 (1971) [Sov. Phys. JETP {\bf 34}, 709 (1972)]; Zh. Eskp. Teor, Fiz. {\bf 63}, 1586 (1972) [Sov. Phys. JETP {\bf 36}, 840 (1973)]; Pis'ma Zh. Eksp. Teor. Fiz. {\bf 18}, 435 (1973) [JETP Lett. {\bf 18}, 255 (1973)].
%
\bibitem{nar-nik_1974} N.~B.~Narozhnyi and A.~I.~Nikishov, Zh. Eksp. Teor. Fiz. {\bf 65}, 862 (1973) [Sov. Phys. JETP {\bf 38}, 427 (1974)].
%
\bibitem{mostepanenko_1974} V.~M.~Mostepanenko and V.~M.~Frolov, Yad. Fiz. {\bf 19}, 885 (1974) [Sov. J. Nucl. Phys. {\bf 19}, 451 (1974)].
%
\bibitem{popov_2001} V.~S.~Popov, Pis'ma Zh. Eksp. Teor. Fiz. {\bf 74}, 151 (2001) [JETP Lett. {\bf 74}, 133 (2001)].
%
\bibitem{avetissian_pre_2002} H.~K.~Avetissian, A.~K.~Avetissian, G.~F.~Mkrtchian, and Kh.~V.~Sedrakian, Phys. Rev.~E {\bf 66}, 016502 (2002).
%
\bibitem{dipiazza_prd_2004} A.~Di Piazza, Phys. Rev. D {\bf 70}, 053013 (2004).
%
\bibitem{bulanov_2006} S.~S.~Bulanov, N.~B.~Narozhny, V.~D.~Mur, and V.~S.~Popov, Zh. Eksp. Teor. Fiz. {\bf 129}, 14 (2006) [J.~Exp. Theor. Phys. {\bf 102}, 9 (2006)].
%
\bibitem{hebenstreit_prl_2009} F.~Hebenstreit, R.~Alkofer, G.~V.~Dunne, and H.~Gies, Phys. Rev. Lett. {\bf 102}, 150404 (2009).
%
\bibitem{ruf_prl_2009} M.~Ruf, G.~R.~Mocken, C.~M\"uller, K.~Z.~Hatsagortsyan, and C.~H.~Keitel, Phys. Rev. Lett. {\bf 102}, 080402 (2009).
%
\bibitem{mocken_pra_2010} G.~R.~Mocken, M.~Ruf,  C.~M\"uller, and C.~H.~Keitel, Phys. Rev.~A {\bf 81}, 022122 (2010).
%
%
\bibitem{dumlu_prd_2010} C.~K.~Dumlu, Phys. Rev.~D {\bf 82}, 045007 (2010).
%
\bibitem{fillion_pra_2012} F.~Fillion-Gourdeau, E.~Lorin, and A.~D.~Bandrauk, Phys. Rev.~A {\bf 86}, 032118 (2012).
%
\bibitem{abdukerim_plb_2013} N.~Abdukerim, Z.~Li, and B.~Xie, Phys. Lett. B {\bf 726}, 820 (2013).
%
\bibitem{li_prd_2015} Z.~L.~Li, D.~Lu, and B.~S.~Xie, Phys. Rev.~D {\bf 92}, 085001 (2015).
%
\bibitem{aleksandrov_prd_2017} I.~A.~Aleksandrov, G.~Plunien, and V.~M.~Shabaev, Phys. Rev. D {\bf 95}, 056013 (2017).
%
\bibitem{woellert_prd_2015} A.~W\"ollert, H.~Bauke, and C.~H.~Keitel, Phys. Rev. D {\bf 91}, 125026 (2015).
%
\bibitem{woellert_plb_2016} A.~W\"ollert, H.~Bauke, and C.~H.~Keitel, Phys. Lett. B {\bf 760}, 552 (2016).
%
\bibitem{aleksandrov_prd_2016} I.~A.~Aleksandrov, G.~Plunien, and V.~M.~Shabaev, Phys. Rev. D {\bf 94}, 065024 (2016).
%
\bibitem{fradkin_gitman_shvartsman} E.~S.~Fradkin, D.~M.~Gitman, and S.~M.~Shvartsman, {\it Quantum Electrodynamics with Unstable Vacuum} (Springer-Verlag, Berlin,~1991).
%
%
\bibitem{dumlu_prl_2010} C.~K.~Dumlu and G.~V.~Dunne, Phys. Rev. Lett. {\bf 104}, 250402 (2010).
%
\bibitem{akkermans_prl_2012} E.~Akkermans and G.~V.~Dunne, Phys. Rev. Lett. {\bf 108}, 030401 (2012).
%
\bibitem{gies_prl_2016} H.~Gies and G.~Torgrimsson, Phys. Rev. Lett. {\bf 116}, 090406 (2016).
%
\bibitem{gies_prd_2017} H.~Gies and G.~Torgrimsson, Phys. Rev.~D {\bf 95}, 016001 (2017).
%
\bibitem{kohlfuerst_prd_2013} C.~Kohlf\"urst, M.~Mitter, G.~von Winckel, F.~Hebenstreit, and R.~Alkofer, Phys. Rev. D {\bf 88}, 045028 (2013).
%
\bibitem{hebenstreit_plb_2014} F.~Hebenstreit and F.~Fillion-Gourdeau, Phys. Lett. B {\bf 739}, 189 (2014).
%
\bibitem{hebenstreit_plb_2016} F.~Hebenstreit, Phys. Lett. B {\bf 753}, 336 (2016).
%
\bibitem{fillion_arxiv_2017} F.~Fillion-Gourdeau,  F.~Hebenstreit, D.~Gagnon, and S.~MacLean, Phys. Rev. D {\bf 96}, 016012 (2017).
%
\bibitem{hebenstreit_prl_2011} F.~Hebenstreit, R.~Alkofer, and H.~Gies, Phys. Rev. Lett. {\bf 107}, 180403 (2011).
%
\bibitem{kohlfuerst_plb_2016} C.~Kohlf\"urst and R.~Alkofer, Phys. Lett.~B {\bf 756}, 371 (2016).
%
\bibitem{schutzhold_prl_2008} R.~Sch\"utzhold, H.~Gies, and G.~Dunne, Phys. Rev. Lett. {\bf 101}, 130404 (2008).
%
\bibitem{akal_prd_2014} I.~Akal, S.~Villalba-Ch\'avez, and C.~M\"uller, Phys. Rev. D {\bf 90}, 113004 (2014).
%
\bibitem{otto_plb_2015} A.~Otto, D.~Seipt, D.~Blaschke, B.~K\"ampfer, and S.~A.~Smolyansky, Phys. Lett. B {\bf 740}, 335 (2015).
%
\bibitem{schneider_jhep_2016} C.~Schneider and R.~Sch\"utzhold, J.~High Energy Phys. 02 (2016) 164.
%
\bibitem{nuriman_plb_2012} A.~Nuriman, B.~Xie, Z.~Li, and D.~Sayipjamal, Phys. Lett. B {\bf 717}, 465 (2012).
%
\bibitem{linder_prd_2015} M.~F.~Linder, C.~Schneider, J.~Sicking, N.~Szpak, and R.~Sch\"utzhold, Phys. Rev.~D {\bf 92}, 085009 (2015).
%
\bibitem{torgrimsson_2017} G.~Torgrimsson, C.~Schneider, J.~Oertel, and R.~Sch\"utzhold, 
J.~High Energy Phys. 06 (2017) 043.
%
\bibitem{allor_prd_2008} D.~Allor, T.~D.~Cohen, and D.~A.~McGady, Phys. Rev. D {\bf 78}, 096009 (2008).
%
\bibitem{gavrilov_prd_2012} S.~P.~Gavrilov, D.~M.~Gitman, and N.~Yokomizo, Phys. Rev. D {\bf 86}, 125022 (2012).
%
\bibitem{fillion_prb_2015} F.~Fillion-Gourdeau and S.~MacLean, Phys. Rev. B {\bf 92}, 035401 (2015).
%
\bibitem{szpak_njp_2012} N.~Szpak and R.~Sch\"utzhold, New J. Phys. {\bf 14}, 035001 (2012).
%
\bibitem{kasper_plb_2016} V.~Kasper, F.~Hebenstreit, M.~Oberthaler, and J.~Berges, Phys. Lett. B {\bf 760}, 742 (2016).
%
\bibitem{kasper_njp_2017} V.~Kasper, F.~Hebenstreit, F.~Jendrzejewski, M.~Oberthaler, and J.~Berges, New J. Phys. {\bf 19}, 023030 (2017).
%
\bibitem{gelis_prl_2006} F.~Gelis, K.~Kajantie, and T.~Lappi, Phys. Rev. Lett. {\bf 96}, 032304 (2006).
%
\bibitem{tanji_aop_2010} N.~Tanji, Ann. Phys. {\bf 325}, 2018 (2010).
%
\bibitem{ruggieri_npa_2015} M.~Ruggieri, S.~Plumari, F.~Scardina, and V.~Greco, Nucl. Phys. {\bf A941}, 201 (2015).


\end{thebibliography}
\end{document}